\documentclass[twocolumngrid,prd,superscriptsize,reprint]{revtex4-1}
\usepackage[]{graphicx}
\usepackage{dcolumn}
\usepackage{bm}

\usepackage{amsmath,amssymb,amsfonts}
\usepackage{fancyhdr}
\usepackage{lipsum} 
\usepackage{subcaption}
\usepackage{multirow}

 \usepackage[colorlinks]{hyperref}
 \hypersetup{
     colorlinks = false,
   }

\newcommand{\comment}[1]{}

\usepackage{color}
\usepackage{babel}
\newcommand{\bea}{\begin{eqnarray}}
\newcommand{\eea}{\end{eqnarray}}

\newcommand{\be}{\begin{equation}}
\newcommand{\ee}{\end{equation}}

\begin{document}

\title[]{Spectral walls in antikink-kink scattering in the $\phi^6$ model}

\author{C. Adam}
\email[]{adam@fpaxp1.usc.es}
\affiliation{Departamento de F\'isica de Part\'iculas, Universidad de
Santiago de Compostela and \\
Instituto Galego de F\'isica de Altas Enerxias (IGFAE), E-15782
Santiago de Compostela, Spain}

\author{K. Oles}
\email[]{katarzyna.slawinska@uj.edu.pl}
\author{T. Romanczukiewicz}
\email[]{tomasz.romanczukiewicz@uj.edu.pl}
\author{A. Wereszczynski}
\email[]{andrzej.wereszczynski@uj.edu.pl}
\affiliation{Institute of Theoretical Physics, Jagiellonian University,
Lojasiewicza 11, Krak\'{o}w, Poland}

\begin{abstract}
We show that thick spectral walls exist in antikink-kink collisions in the $\phi^6$ model. In this model, they are triggered by the so-called {\it delocalized modes} which do not exist in the single-soliton sector but emerge in antikink-kink ($\bar{K}K$) collisions. Therefore, spectral walls are a rather common phenomenon that should occur in many solitonic collisions involving non-symmetric kinks as, e.g., in the $\phi^8$ or higher power models.
\end{abstract}
\maketitle

%%%%%%%%%%%%%%%%%%%%%%%%%%%%%
\section{Introduction}
%%%%%%%%%%%%%%%%%%%%%%%%%%%%%

Although they are the simplest topological solitons \cite{R, SM, Shnir}, kinks can reveal a complicated and fascinating pattern of interactions. There are several reasons for this. First of all, kinks comprise two rather distinct features. They are {\it localized} solutions of nonlinear field equations in (1+1) dimensions which are characterized by a {\it global} quantity, that is, a topological charge. Secondly, there are various ways in which kinks can interact. Except for the rare Bogomol’nyi-Prasad-Sommerfield (BPS) cases \cite{Bo, JT, Ma8}, kinks experience a static force if placed at a finite distance from each other. Further, as a localized particle like object, a kink can possess internal degrees of freedom (DoF) usually identified with normal (or quasi-normal) modes arising in the standard linear perturbation theory. This can be further generalized to {\it delocalized modes} or delocalized DoF, which exist {\it only} for multi-kink configurations. There is also radiation (scattering modes) which may have a very nontrivial impact on kinks like, e.g., negative radiation pressure \cite{neg-rad}. 

In particular, the interplay between the kinetic motion of kinks and their internal modes (both localized and non-localized ones), can have a very nontrivial impact on kink dynamics. 
The first famous example is the {\it resonant energy transfer} \cite{Sug, CSW}, which explains the appearance of a fractal structure in the final state formation in kink-antikink collisions in various models like, e.g., $\phi^4$ theory \cite{Sug, CSW, MORW}. Let us consider a kink and an antikink which initially carry only kinetic energy since they are simply boosted towards each other. During the collision the energy can be transferred to internal DoF (e.g., the shape mode or Derrick modes for kinks in the $\phi^4$ model). As a result, it may happen that the kink and antikink possess too little kinetic energy to overcome the attractive static force. This results in kink-antikink annihilation by the formation of a so-called {\it bion} which decays by the emission of radiation. However, the energy can also be transferred back to the kinetic DoF which can allow the solitons to escape. The actual final state (annihilation or back scattering) depends very sensitively on the initial velocity and reveals a fractal-like structure. It is important to remark that the internal modes participating in the resonant energy transfer can be localized on the constituent solitons (as in the $\phi^4$ model \cite{Sug, CSW, Weigel, MORW, AMORW}) or can be delocalized between the colliding kinks (as happens in antikink-kink collisions in $\phi^6$ theory \cite{Tr, phi-6}).

The second phenomenon which involves kinetic and internal DoF is the {\it spectral wall phenomenon} \cite{spectral-wall}. A spectral wall is an unstable stationary solution caused by a normal mode crossing the mass threshold, i.e., by the transition of a normal mode frequency into the continuous spectrum. It results, for example, in a kink-antikink pair frozen at a certain mutual distance $2a_{sw}$, where the constituent solitons are subject to small oscillations. Equivalently, a spectral wall can be viewed as a barrier in soliton dynamics located at the point where this unstable solution can be formed. 

Specifically, if we collide a kink and an antikink carrying an exited internal mode which for a certain intersoliton distance crosses the mass threshold, then three possible scenarios can be observed. If the amplitude $A$ of the mode is equal to a critical value $A=A_{cr}$, then the incoming kink and antikink form this stationary, (arbitrarily) long living solution with their average positions frozen at a certain value $\pm a_{sw}$, which is identified with the position of the spectral wall. It has been verified that the position of the spectral wall is a universal observable and does not depend on the particularities of the colliding solitons (e.g., their velocity). It is, in fact, uniquely determined by the inter-soliton distance at which the mode crosses the mass threshold. For $A<A_{cr}$, the incoming solitons pass through the spectral wall and their dynamics is less and less affected by the wall as $A$ decreases. If the mode is strongly excited, $A>A_{cr}$, then the solitons are reflected before the spectral wall, and the reflection distance grows with the amplitude of the mode. Such a spectral wall barrier may exist even when the solitons are far away from each other. Thus, a spectral wall can influence also the long range intersoliton interactions.  

Qualitatively, this behavior can be understood by the observation that the normal mode carries a certain fraction of the energy of the configuration which grows with the amplitude. If it carries too much energy, then the system is not capable of transferring this energy to other DoF before the mode disappears into the continuum, and the wall crossing cannot occur.

Importantly, a spectral wall is a very selective phenomenon. Every mode which crosses the mass threshold has its own spectral wall. Furthermore, the excitation of other modes does not affect the transition/reflection at a particular spectral wall. Only the amplitude of the mode related to the spectral wall does matter. 

Originally, spectral walls have been observed in BPS collisions, i.e., in models which allow for static solutions describing a kink-impurity or kink-antikink pair at any distance \cite{sw-1, 2field-SW}.  Such BPS collisions, although important, are rather rare and require either the addition of a background field (an impurity) \cite{BPS_imp-1, solv} or some special multi-field theories where the coupling constants are fine-tuned to support a nontrivial BPS sector \cite{Bazeia, Izq-1, Izq-2}.  

Recently, spectral walls have also been found in near-BPS collisions, where a small kink-antikink force exists \cite{thick-SW}. Strictly speaking, in this case the spectral wall transmutes into a {\it thick spectral wall}, where stationary solutions are located at a larger distance than in the BPS version $a_{sw}^{thick} > a_{sw}$. This is related to the fact that the stationary state is formed {\it before} the mode crosses the mass threshold. In this case, the formation of a thick spectral wall is due to a balance between the kink-antikink attractive force (which would be absent in the BPS case) and a repulsive interaction with the mode. For a detailed mathematical explanation we refer to sec. V in \cite{thick-SW}.  As a consequence, the position of the stationary solution is not uniquely defined for a given mode but now depends on the velocity of the scattered kinks. Equivalently, if we prepare as an initial state a pair of kink and antikink with a separation of  $2a>2a_{sw}$ and with the pertinent mode exited, then there is a critical value of the amplitude of the mode for which such a stationary state is formed already at this separation. Furthermore, the value of the amplitude depends on the initial separation. When the separation approaches the position of the original spectral wall in the BPS limit, $a \to a_{sw}$, the critical amplitude wich generates the stationary solution diverges. Hence, the thick spectral wall is bounded by the usual (BPS) spectral wall. Nonetheless, a thick spectral wall keeps its selective nature. 
 
A common property of the (thick) spectral walls is that they require rather slow velocities of the colliding solitons. This is due to the fact that the critical amplitude for which the stationary solution is formed grows with the velocity. Thus, at some point nonlinear effects begin to dominate the linear mode picture. Furthermore, for high speed collisions the system evolves too quickly to allow for the existence of spectral walls. In a sense, in such a rapid, non-adiabatic evolution the solitons do not have time to realize that a mode disappeared into the continuum.  This is precisely the reason why there are no spectral walls in kink-antikink collisions in the $\phi^4$ model, see for example \cite{Izq-3} where kinks with excited shape mode have been studied. Here, the initial states, which are free kinks, possess a normal mode (shape mode) which necessarily temporarily disappears during the collision. Indeed, the vacuum, which is a configuration realized at a certain instant in the scattering, does not support any modes. Similarly, no spectral walls have been found in the double sine-Gordon model \cite{Azadeh}. 

We can conclude that (thick) spectral wall will not always exist in kink-antikink collisions, even if the colliding solitons possess normal modes localized on the solitons. In this situation, the modes cross the threshold when the solitons are very close to each other entering into a very rapid (non-adiabatic) phase of the process. To find spectral walls in usual, non-BPS, kink scattering processes, we need modes which cross the mass threshold when the inter-soliton separation is still large and the system undergoes an adiabatic evolution. 

It is the aim of the current paper to show that thick spectral walls can easily exist in collisions of non-symmetric kinks and antikinks. In this case, there are arbitrarily many {\em delocalized} modes trapped between the colliding solitons. These modes, one by one, cross the mass threshold as the kinks approach each other. Since this can happen at a very large separation, we find a clear evidence for spectral walls. 

%%%%%%%%%%%%%%%%%%%%%%%%%%%%%
\section{Spectral walls in the $\phi^6$ model}
%%%%%%%%%%%%%%%%%%%%%%%%%%%%%
Let us consider the simplest model with non-symmetric kinks, that is, $\phi^6$ theory
\be
L_{\phi^6}[\phi]= \int_{-\infty}^\infty \left( \frac{1 }{2} \phi_t^2 - \frac{1}{2}\phi_x^2 - \frac{1}{2} \phi^2\left(1-\phi^2\right)^2 \right) dx.
\ee
This model supports a static kink
\be
\Phi_{K} (x;a)\equiv \phi_{(0,1)}(x;a) =\sqrt{\frac{1+\tanh(x-a)}{2}}
\ee
which joins two vacua: $\Phi_K(x=-\infty)=0$ and $\Phi_K(x=+\infty)=1$. The antikink interpolates between  the same vacua but in opposite order 
\be
\Phi_{\bar{K}} (x;a) \equiv \phi_{(1,0)}(x;a) =\sqrt{\frac{1-\tanh(x-a)}{2}}
\ee
Obviously, these solitons are not anti-symmetric $\Phi_K(x) \neq -\Phi_{\bar{K}}(-x)$. 
 \begin{figure}
 \includegraphics[width=1.00\columnwidth]{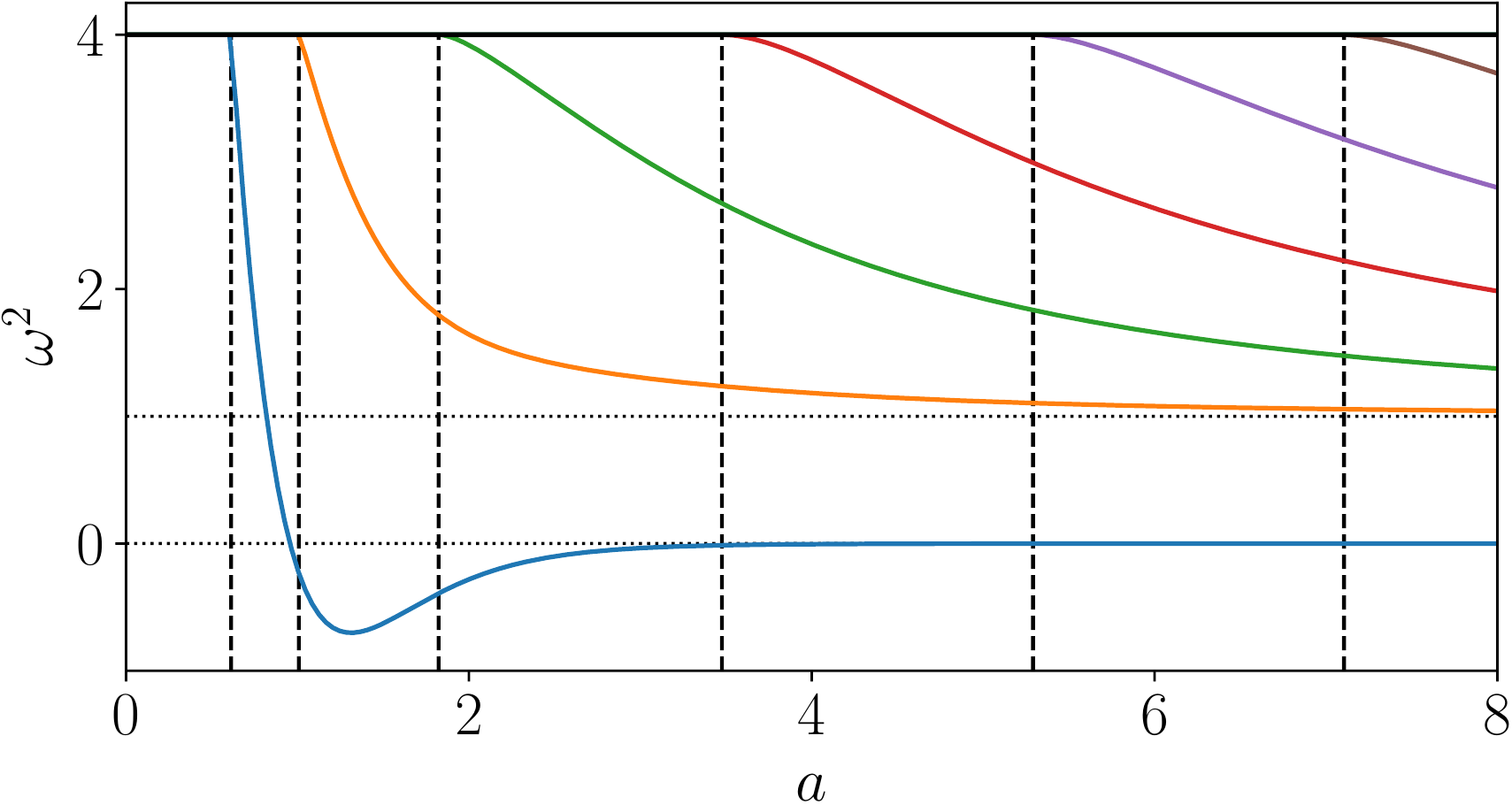}
 \caption{Frequency of the delocalized antikink-kink even modes as a function of the distance $a$. Vertical lines denote the positions of the spectral walls, i.e., the points where the modes hit the mass threshold.}\label{modes}
 \end{figure}

Furthermore, a single kink (and antikink) does not support any discrete normal modes except for the usual zero mode. Importantly, due to a different curvature of the field theoretical potential at the vacua, the effective potential in the linear perturbation problem has two {\it different} mass thresholds. Namely, at the vacuum $\phi=0$ the mass of small perturbations is $m_0=1$ while at the vacuum $\phi=1$ we find $m_1=2$. This has a huge impact on the collisions of these solitons. For simplicity we call such solitons non-symmetric kinks.

In kink-antikink collisions the solitons are joined by the $\phi=1$ vacuum which effectively produces a potential barrier of hight $m_1^2=4$ between the two potential wells which are generated by single soliton solutions. Therefore, the spectral structure for this sector is just a sum of two single soliton sectors and no massive normal modes show up.

However, in the antikink-kink sector, where the solitons are connected by the $\phi=0$ vacuum something new happens. The single-soliton potential wells are joined by the lower mass threshold, effectively forming a much larger antikink-kink potential well which can support new, so-called {\it delocalized} modes, which are trapped between the colliding solitons. Their number depends of the antikink-kink separation $2a$ and increases arbitrarily as the separation grows. To see this, we consider the usual linear perturbation theory where we perturb a configuration which is a simple sum of an antikink and a kink located at $-a$ and $a$, respectively. This leads to the following spectral problem
\be
\left. \left( \frac{d^2}{dx^2}  - \frac{d^2U}{d\phi^2}\right|_{\Phi_{\bar{K}}(x;-a)+\Phi_K(x;a)} \right) \eta(x;a) = - \omega^2\eta (x;a),
\label{spectral}
\ee 
Here $U=\frac{1}{2}\phi^2(1-\phi^2)^2$ is just the $\phi^6$ potential. 
Then, the form of the normalized modes $\eta$ as well as their frequency $\omega$ and number depend on the separation, see  Fig. \ref{modes}, where the first few even modes are shown. They cross the mass threshold at $a=1.0072$ ($n=1$ mode), $a=1.8229$ ($n=2$ mode), $a=3.4763$ ($n=3$ mode), $a=5.2910$ ($n=4$ mode), $a=7.1050$ $(n=5$ mode). There is also an unstable mode with imaginary frequency. This reflects the fact that the simple sum $\Phi_{\bar{K}}(x;-a)+\Phi_K(x;a)$ is not a static solution of the model, because the antikink and kink attract each other. Nonetheless, for larger $a$, this unstable mode is almost zero as the static force decreases exponentially. Therefore, for a sufficiently large distance $a$ the naive sum can be treated as a solution with a good approximation. 

These delocalized two-soliton modes are responsible for the fractal structure in the final state formation. This was originally found in \cite{Tr} and recently carefully further investigated in \cite{phi-6}.

\begin{figure}
\hspace*{-0.3cm}  \includegraphics[width=1.10\columnwidth]{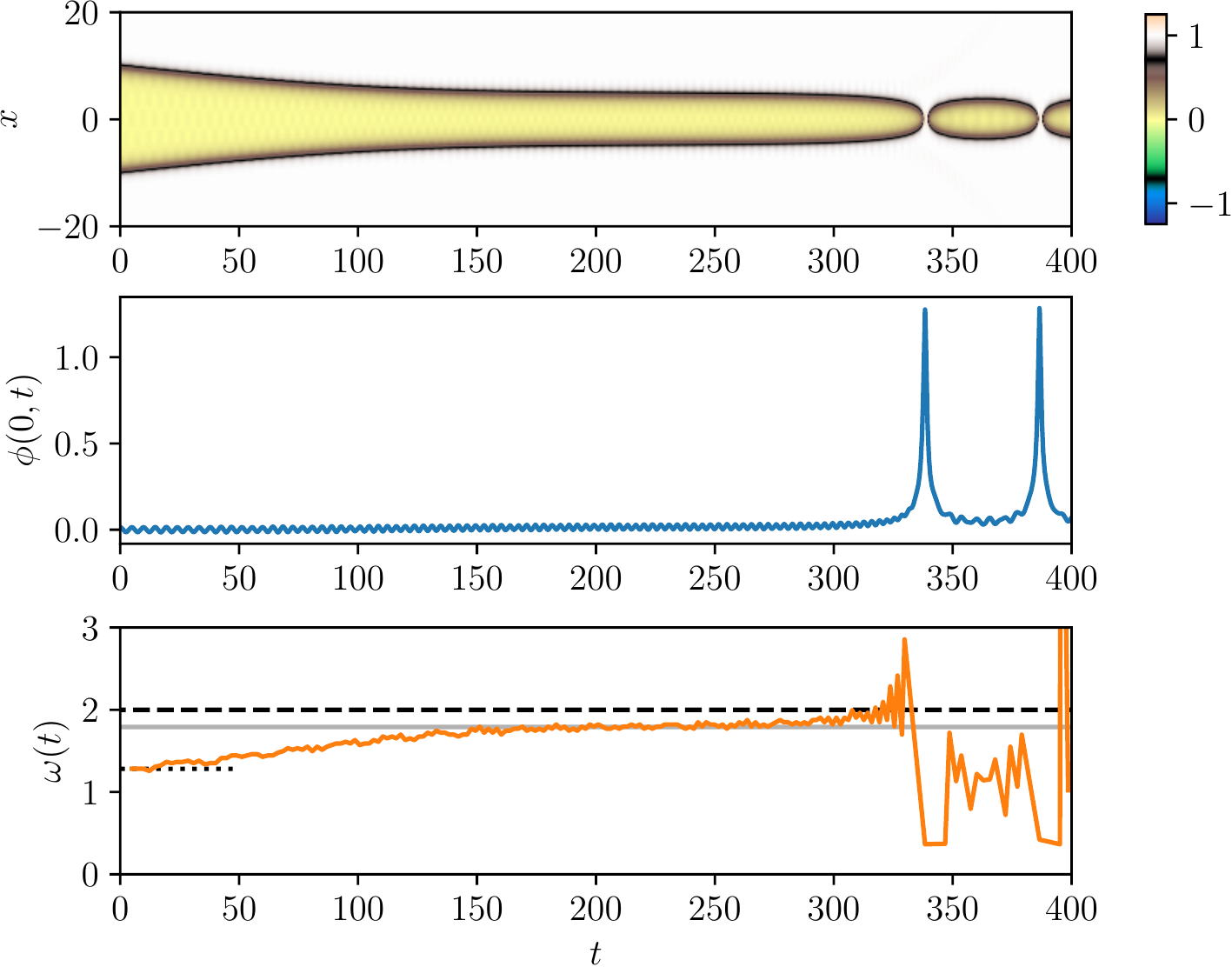}
 \caption{Dynamics of an antikink-kink pair  with the fourth delocalized mode excited. See description in the text.}\label{evolution}
 \end{figure}

Here, however, it is important that the delocalized modes can hit the mass threshold (of the antikink-kink collision) at an arbitrarily large distance, where the collision is still in a slow velocity (adiabatic) phase. These are ideal conditions for the appearance of spectral walls. 

 In our first numerical experiment, we take as an initial configuration a pair of antikink and kink which are located at $\mp a=10$, respectively, are boosted toward each other with $v_{in}=0.05$. Furthermore, we add  the fourth delocalized mode (purple mode in Fig. \ref{modes}) $\eta_4$ with frequency $\omega_4$ and initial amplitude $A_4=0.0464$
 \bea
  \phi_{in}(x,t) &=&  \sqrt{\frac{1-\tanh \gamma (x - v_{in}t +a)}{2}} \label{pert-mode}  \\ 
&& +\sqrt{\frac{1+\tanh \gamma (x + v_{in} t-a)}{2}} + A_4 e^{i\omega_4 t} \eta_4 (x;a) \nonumber
 \eea
 where $\gamma =(1-v_{in}^2)^{-1/2}$.
In Fig. \ref{evolution} we show the evolution of such a colliding initial configuration. At the beginning, due to the non-zero initial velocity, the solitons approach each other but quite quickly stabilize and form a long living stationary state, where the positions of the kink and antikink are frozen, see top panel in  Fig. \ref{evolution}. This is a thick spectral wall or, strictly speaking, one of the stationary solutions forming the thick spectral wall related to the first delocalized mode. After a long time, the amplitude of the mode decreases due to the energy transfer to radiation. As a consequence, the stationary solution destabilizes and the solitons can pass through the obstacle. They collide, performing in this case a few bounces. After the first collision the biggest part of the energy is transferred from the initially exited fourth mode into the first mode. In the lower panel we show the time dependence of the value of the field at the origin $\phi(0,t)$. During the scattering the field oscillates because of the excitation of the fourth delocalized mode. In the bottom panel, we present the frequency of the delocalized mode while the antikink and kink approach each other and form the thick spectral wall. The frequency is computed from the actual field oscillations at the origin. As follows from the two-soliton linear perturbation theory, the frequency grows with decreasing $a$. The dependence $\omega(a)$ is very well approximated by the linear perturbation theory, eq. (\ref{spectral}). 

\begin{figure}
 \includegraphics[width=1.00\columnwidth]{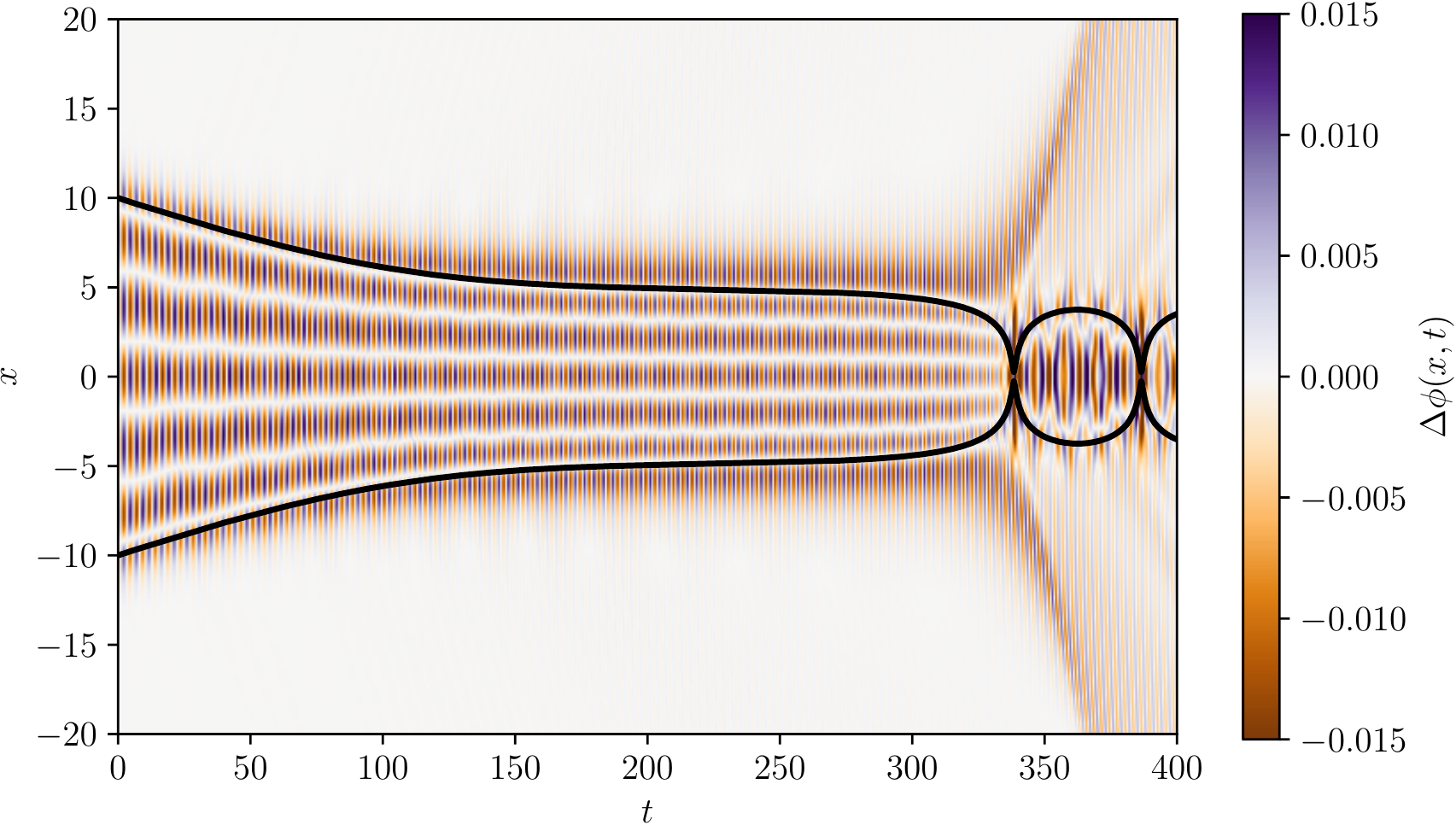}
 \caption{Dynamics of the same state as in Fig. \ref{evolution}. We plot the time evolution of the field with the subtraction of the single soliton profiles: $\Delta \phi(x,t)= \phi(x,t) - \Phi_{\bar{K}}(x;-a)-\Phi_K(x;a)$. The black line denotes the position of the solitons.}\label{evolution-mode}
 \end{figure}

In Fig. \ref{evolution-mode} we consider the same collision and present the field with the subtraction of the naive superposition $\phi(x,t)-\Phi_{\bar{K}}(x;-a)-\Phi_K(x;a)$, which allows us to clearly visualize the excited mode. 
 
Similar numerical computations are performed for the first four lowest lying modes $\eta_n$ (orange, green, red and purple modes in Fig. \ref{modes}) and several initial separations  $2a$ between the solitons. For reasons of simplicity, we use a slightly different strategy to find the corresponding thick spectral walls. Namely, we choose the initial configuration
\bea
  \phi_{in}(x,t) &=& \sqrt{\frac{1-\tanh(x+a)}{2}} \label{pert-modes}  \\ 
&& +\sqrt{\frac{1+\tanh(x-a)}{2}} + A_n e^{i\omega_n t} \eta_n (x;a), \nonumber
 \eea
 which is a non-boosted antikink-kink pair with the addition of one of the first few (lowest frequency) modes $\eta_n$ with frequency $\omega_n$ and initial amplitude $A_n$ (no summation over the index $n$ is assumed). The addition of a delocalized mode provides a repulsion acting against the attractive intersoliton force, which for a suitably chosen amplitude $A_n$  of the $n$-th mode leads to the formation of a stationary state, i.e., a thick spectral wall. For a given $n$ and $a$, we search for a critical value of the amplitude $A_n$ such that the stationary configuration (\ref{pert-modes}) is a solution for all times. We repeat this procedure for various initial positions $a$ and for $n=1,2,3,4$. It turns out that this strategy allows for a more efficient detection of the thick spectral walls.
 
\begin{figure}
 \includegraphics[width=1.00\columnwidth]{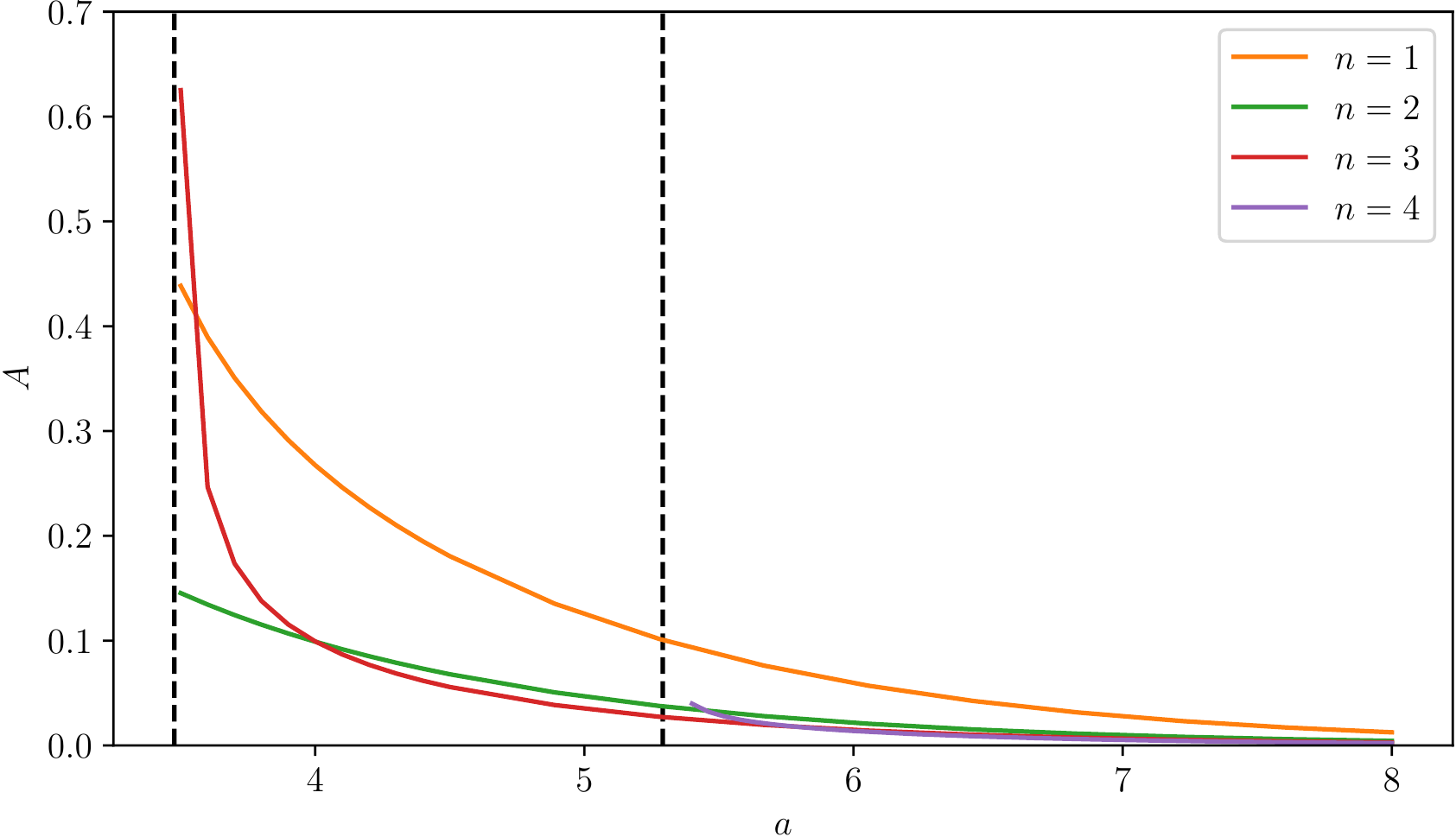}
 \caption{Amplitude of the first four delocalized normal modes for which a thick spectral wall is formed, as a function of the antikink and kink positions $\mp a$. The dashed vertical lines denote the positions of the solitons at which the third and fourth mode hit the mass threshold. }\label{amplitude}
 \end{figure}

 The results are summarized in  Fig. \ref{amplitude} where we plot the critical amplitude of the mode $A_n$ for which the stationary solution (the thick spectral wall) is formed as a function of the position of the antikink and kink, $\mp a$. 

 For the third and fourth positive frequency delocalized mode, we clearly see that the position of the thick spectral wall is bounded from below by the point where the mode hits the mass threshold, as predicted from the linear perturbation analysis, eq. (\ref{spectral}) (see also Fig. \ref{modes}). Indeed, if the initial separation tends to the separation at which the mode enters the continuum, $2a\to 2a_{sw}$, then the value of the critical amplitude for which the thick spectral wall is formed diverges. 
 
For the first two modes, $a_{sw}$ is located at much smaller values, which requires very huge amplitudes to stabilize the antikink-kink pair, even if we are significantly before the position of these spectral walls. This of course means that nonlinearities begin to play an important role and the stationary solutions cease to exist. 

%%%%%%%%%%%%%%%%%%%%%%%%%%%%%
\section{Conclusions}
%%%%%%%%%%%%%%%%%%%%%%%%%%%%%

In the present work, we demonstrated the existence of thick spectral walls in antikink-kink scattering in the $\phi^6$ model. They arise due to the existence of delocalized (two-soliton) normal modes which can cross the mass threshold at a large distance, which guarantees that the dynamics has an adiabatic-like character. 

Such thick spectral walls are stationary solutions formed by a colliding antikink and kink, where the constituents solitons freeze at a certain mutual distance, performing small oscillations. The actual positions of these stationary solutions can be predicted from perturbation theory, where the attractive inter-soliton force is balanced by a repulsion due to the excitation of the mode, see \cite{thick-SW}. 

The observed thick spectral walls act as obstacles in antikink-kink collisions and, therefore, have a nontrivial effect on their dynamics. 
On the other hand, there are no spectral walls in kink-antikink collisions in the $\phi^6$ model, because there are no delocalized modes in this sector. 

We expect that similar thick spectral walls will be present in antikink-kink collisions in any field theory provided that {\it (i)}  the single kink and antikink are non-symmetric solitons with two different masses of small perturbations at the vacua, i.e., two different mass thresholds; {\it (ii)} the scattering kinks are joined by the lower mass vacuum (lower mass threshold). This guarantees the emergence of delocalized, trapped modes which necessarily should lead to the appearance of thick spectral walls. Thus, they should be visible, non only in versions of the $\phi^6$ model, e.g., \cite{new-phi6, simas-new, takyi}, but also in antikink-kink collisions in the $\phi^8$ model. In fact, the observed repulsion in the latter model, which occurs for an initial configuration which is a naive superposition of an antikink and a kink \cite{GB}, is probably an effect produced precisely by these thick spectral walls. The same should happen for other models with so-called fat tails \cite{higher-nick, kev, Decker-1, Mohammadi-2, khare}. 

To conclude, thick spectral walls are a generic phenomenon influencing kink dynamics in many field theoretical systems. 
%%%%%%%%%%%%%%%%%%%%%%%%%%%%%%%%%%%%
\section*{Acknowledgements}
%%%%%%%%%%%%%%%%%%%%%%%%%%%%%%%%%%%%

C. A. and A. W. 
acknowledge financial support from the Ministry of Education, Culture, and Sports, Spain (Grant No. PID2020-119632GB-I00), the Xunta de Galicia (Grant No. INCITE09.296.035PR and Centro singular de investigación de Galicia accreditation 2019-2022), the Spanish Consolider-Ingenio 2010 Programme CPAN (CSD2007-00042), and the European Union ERDF.

K. O., T. R., and A. W. were supported by the Polish National Science Centre (Grant No. NCN 2019/35/B/ST2/00059).

%\newpage

\end{document}